# Full field chemical imaging of buried native sub-oxide layers on doped silicon patterns

F. de la Peña [a], N. Barrett [b],  L.F. Zagonel [a], M. Walls [a], O. Renault [c]

[a] Univ Paris-Sud, Laboratoire de Physique des Solides, CNRS/UMR 8502, Orsay, F-91405, France
[b] CEA, IRAMIS, SPCSI, LENSIS, F-91191 Gif-sur-Yvette, France
[c] CEA-LETI, MINATEC, 17 rue des Martyrs, 38054 Grenoble Cedex 9, France

a b s t r a c t

Fully energy-filtered X-ray photoelectron emission microscopy is used to analyze the spatial distribution of the silicon sub-oxide structure at the $SiO_2$/Si interface as a function of underlying doping pattern. Using a spectroscopic pixel-by-pixel curve fitting analysis, we obtain the sub-oxide binding energy and intensity distributions over the full field of view. Binding energy maps for each oxidation state are obtained with a spatial resolution of 120 nm. Within the framework of a five-layer model, the experimental data are used to obtain quantitative maps of the sub-oxide layer thickness and also their spatial distribution over the p–n junctions. Variations in the sub-oxide thicknesses are found to be linked to the level and type of doping. The procedure, which takes into account instrumental artefacts, enables the quantitative analysis of the full 3D dataset.

## 1. Introduction

In silicon based micro- and nanotechnologies, the control of the quality and thickness of the $SiO_2$/Si interface is of prime importance. An interface of several atomic layers, with a high defect concentration can result in fixed charges and contribute significantly to variations in electrical properties. One of the limits on CMOS downscaling is the quality of the interface. The chemistry of the $SiO_2$/Si interface has already been studied in the laboratory using both angle-resolved X-ray photoelectron spectroscopy (XPS) with Al Kα or Mg Kα X-ray sources and synchrotron radiation-based photoelectron spectroscopy (PES). The use of XPS to obtain accurate values for the thickness of the $SiO_2$ layer and the sub-oxide interfaces has attained extremely high standards in order to define a universal method independent of specific laboratory conditions [1]. The precision now possible in measuring oxide layer thickness is smaller than typical variations in silicon oxygen bond lengths. Photoelectron spectroscopy is also non-destructive, allowing complementary analyses such as Secondary ion mass spectrometry (SIMS) to be performed a posteriori. Synchrotron radiation is particularly useful for a detailed analysis of the $SiO_2$/Si interface since the depth probed can be changed by tuning the photon energy. Valence band PES gives information on the density of states and the valence band offsets, in particular on the presence of interface dipoles [2,3]. Finally, the width of the photoelectron spectrum may be measured and thus the work function deduced. The original reference for synchrotron radiation-based PES studies of this system is the work of the Himpsel group on Si(100) and Si(111) [4]. Other experimental studies have been carried out by [5,6]. Several models of the interface have been proposed in the light of theoretical calculations [7–10]. However, typical X-ray spot sizes are of the order of a fraction of a millimetre, therefore virtually no spatial information is available. Thus, PES is an area-averaged technique and hence cannot address gate oxide dimensions compatible with real circuits. Downscaling in microelectronics makes the availability of reliable spatially resolved quantification of the interface chemical states a pressing requirement.

Energy resolved X-ray photoelectron emission microscopy (XPEEM) combines the required spatial and chemical state resolution. In particular, by using soft X-ray excitation combined with a full energy-filtered analysis, elemental, chemical and electronic structure sensitivities of classical PES with spatial resolutions on the 100 nm scale are now readily accessible with current PEEM instruments. Image series as a function of photoelectron kinetic energy $E_k$ are acquired step by step over the energy window of interest. Each image gives the intensity as a function of position within the microscope field of view (FoV). A full image series is a 3D dataset, called a Spectrum Image (SI), allowing extraction of photoemission spectra within the FoV from any area of interest (AOI) down to the pixel size. This technique was first applied in a study of silicon anodic oxidation [11], in which the effect

Corresponding author. Tel.: +33 169083272; fax: +33 169088446. E-mail address: nick.barrett@cea.fr (N. Barrett).

## 2. Experiment

The samples, made in the Laboratoire d'Electronique et de Technologie de l'Information (LETI) at MINATEC, CEA-Grenoble, consisted of two-dimensional doped patterns implanted into silicon in the form of lines with variable spacing and widths from 100μm to 0.1 μm [12]. Before implantation, the silicon wafers were clean and had any surface oxide removed by standard chemical procedures. Each set of lines was identified by figures with the same doping level, giving suitable shapes for optimization of the electron optics of the PEEM instrument. Two samples were studied, Fig. 1. Phosphor and boron ions were used for the n- and p-type doping. The first sample, designated $N^+/P^-$, had $N^+$ (representing very high n-type doping) doped zones ($10^{20}$ cm$^{-3}$) implanted into a $P^-$ substrate ($10^{16}$ cm$^{-3}$). The second, designated $P^+/N$, had $P^+$ (representing very high p-type doping) patterns ($10^{20}$ cm$^{-3}$) implanted into an N type ($10^{17}$ cm$^{-3}$) substrate; the latter was itself deposited on a Si $P^-$ wafer ($10^{16}$ cm$^{-3}$). A native oxide layer was present on both samples. Fig. 1 also illustrates schematically the expected band line-ups in such samples. The two samples are not symmetric in their preparation since in both cases the starting substrate is $P^-$. Hence, the $P^+/N$ sample underwent a supplementary ion implantation to obtain the required doping patterns. The doping levels have been measured using secondary ion mass spectrometry (SIMS); the results are given in Table 1.

The spectromicroscopy experiments used a NanoESCA XPEEM system (Omicron Nanotechnology), more fully described elsewhere [16,18]. The apparatus was installed on the CIPO beamline of the ELETTRA synchrotron (Trieste, Italy). The incident photon energy was 127 eV, representing the best compromise between the beamline undulator response and surface sensitivity. At this energy, the inelastic mean free path (IMFP) of the Si 2p electrons is low, thus the substrate signal is expected to reflect bent bands at the substrate/oxide interface rather than the flat band situation. Given the depth sensitivity, photoelectron diffraction contributions from the substrate are also negligible. The photoelectrons are collected within a 6.1° cone around the surface normal. The PEEM contrast aperture diameter was 150 μm, a trade-off between lateral resolution and PEEM column transmission. The double of electrostatic charge on the Si 2p emission due to the anodic oxide was demonstrated. Using this method, a preliminary analysis of the present system revealed doping-dependent variations in the overall oxide layer thickness [12]. However, depending on the size of the AOIs up to 99% of the information in the dataset is not used.

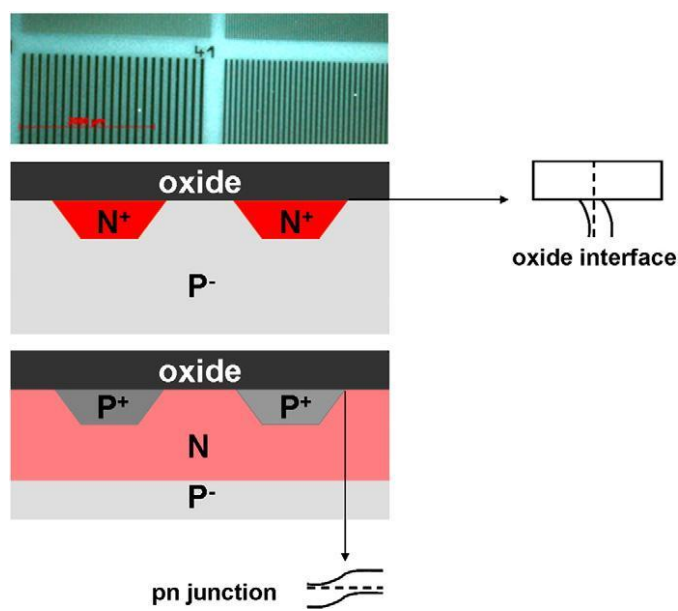

Fig. 1. Optical micrograph of doped pattering schematic cross-section of samples $N^+/P^-$ and $P^+/N$. The oxide/substrate and p–n band line-ups are also shown.

The most information is available from pixel-by-pixel curve fitting to spectra over the full field. This has already been used in XPEEM analysis to obtain, for example, concentration maps of InAs/GaAs quantum dots and rings [13,14]. Ga 3d and In 4d core levels, separated by 2 eV were resolved allowing elemental mapping of the surface. An alternative full field approach was developed by Ratto et al. [15], however, it relies on the existence of an internal reference in the sample with known composition, which is not always the case. Furthermore, the information extracted is averaged over the photoelectron escape depth. Such analysis is often, but not always, limited to elemental mapping based on well separated core levels. To accurately map the surface and sub-surface spatial variations in the chemical states and binding energies, careful handling of the dataset and flexible numerical analysis tools are required. One must correct for the microscope transmission function (or illumination) at constant kinetic energy in the FoV, the variations in the photon flux and the non-isochromaticity over the FoV [16,17]. When one does this, the inevitably better statistics gives new quantitative information on the spatial distribution of the chemical states and on sub-surface interfaces. As a test case, we have studied the interface structure of the native oxide as a function of the Si(100) substrate doping. The area-averaged Si 2p core-level studies have already shown that the interface structure is strongly dependent on the particular oxide growth conditions [4,5]. We obtain maps of the $SiO_2$ and the sub-oxide distributions. We show that the $SiO_2$/Si interface is not the same over p- and n-type doped patterns, and develop a model to quantify the spatial extent of the different interface sub-oxides between $SiO_2$ and Si.

Table 1
Doping levels as measured by secondary ion mass spectrometry.

| Sample | n-type (at/cm³) | p-type (at/cm³) |
|---|---|---|
| N+/P− | $10^{20}$ | $1\times10^{15}$ |
| P+/N | $2.5\times10^{16}$ | $10^{20}$ |

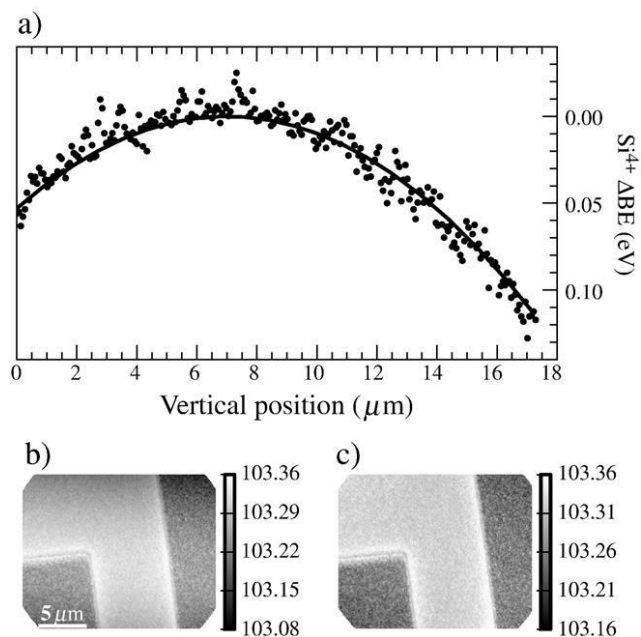

Fig. 2. (a) Non-isochromaticity measurement by a parabolic fit to the $Si^{4+}$ peak energy as a function of the vertical (energy dispersive plane) pixel position in a region far from p–n junctions for sample $N^+/P^-$. (b) $Si^{4+}$ binding energy map obtained from a preliminary fit to the Si 2p spectra of sample $N^+/P^-$ to a model without correction for the non-isochromaticity and (c) the same fit corrected for non-isochromaticity. In (b) and (c) the heavily doped $N^+$ patterns are the bright inverted "L" regions.

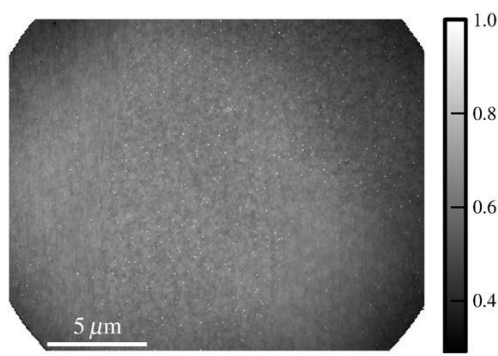

Fig. 3. Normalized map of the number of counts of the final model fit at the highest kinetic energy showing no feature which can be associated with the sample structure. This signal is used as the microscope transmission function.

hemispherical analyser pass energy of the NanoESCA was 100 eV and 1 mm entrance and exit slits were used. Image series through the Si 2p core level were acquired in 0.1 eV steps. The overall estimated energy resolution was 0.42 eV, with a spatial resolution of 120 nm. All results were taken using a 25 μm FoV. In the field of view the heavily doped patterns are the inverted "L" shapes. With 4×4 camera binning, the resulting SI has 320×256 pixels and 81 energy channels. The acquisition time was 3 min per image. The photoelectron energy E is measured with respect to the sample Fermi level $E_F$, thus $E-E_F = E_K + \Phi_{WF}$, where $E_K$ is the photoelectron kinetic energy and $\Phi_{WF}$ the analyser work function. The maximum intrinsic energy dispersion or non-isochromaticity in the vertical direction of the FoV is less than 150 meV (see below). The binding energy scale was calibrated using the Fermi level and the 3d emission of a flat polycrystalline in situ sputtered Ag sample with the same experimental parameters.

## 3. Data analysis

Although each spatial pixel of the SI obtained by Scanning photoelectron microscopy (SPEM) or PEEM contains a standard photoemission spectrum, a quantitative analysis of the full SI involves new challenges. Robust automatic analysis procedures are necessary because there are typically $10^4$ spectra. Principal Components Analysis (PCA) [19–21] and blind source separation [22,23] are very promising because they perform the analysis of the full SI with a minimum of a priori knowledge. However, in their standard form, these techniques assume that the dataset can be represented by a linear model; this is not the case here, since energy shifts are expected as a consequence of the doping level across the FoV.

Another approach is to adapt a curve fitting method proposed by Himpsel [4] to the SI. A naive attempt to perform a standard analysis of each spectrum will fail in general due to the following reasons. First, in the kinetic energy range used, the secondary electron emission (SEE) background is non-negligible and cannot be easily accounted for [24]. Second, the lower signal to noise ratio (SNR) of each pixel (with respect to standard PES experiments or AOI spectra) leads to stability problems in any fitting procedure, particularly if too many free parameters are used. Third, in the presence of a non-negligible SEE background, the standard iterative Shirley background removal cannot be used. Even if the SEE background was negligible or if it was removed by other means, the iterative Shirley background cannot be used with low SNR [24]. Finally, artefacts such as the non-isochromaticity aberration, the inhomogeneous illumination (or transmission) over the FoV and the synchrotron photon flux decay are all significant for quantitative analysis and must be carefully taken into account.

Thus, for accurate results, a curve fitting approach must include careful background subtraction and artefact correction. In addition, special attention must be given to minimize the number of free parameters in the model and to provide sensible starting values for each pixel. We have defined a six component model: the SEE background, approximated by an exponential function of the form $Ae^{-E/\tau}$, and five core components to model the core-level emission from metallic silicon and its four oxidation states $Si^{1+}$, $Si^{2+}$, $Si^{3+}$ and $Si^{4+}$. Each core component is comprised of the sum of two Voigt curves with a 2p spin-orbit splitting of 0.61 eV and a branching ratio of 2:1 [4]. The integral of these curves, weighted by a multiplicative factor, s, is added to account for the Shirley background [25,26]. This procedure assures an accurate background removal which is essential for quantitative determination of sub-oxides [1]. To correct for the photon flux decay (about 15%), the beam current was monitored continuously during the experiment and the intensity of the whole SI was scaled appropriately.

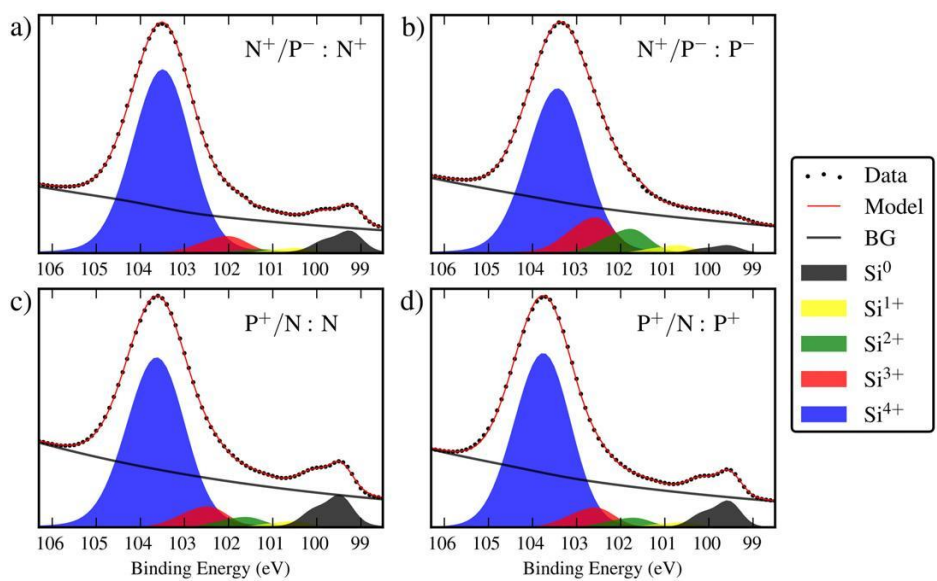

Fig. 4. Full field best fit to the Si 2p spectra averaged over all pixels in a single pattern: (a) $N^+$ of sample $N^+/P^-$; (b) $P^-$ of sample $N^+/P^-$; (c) N of sample $P^+/N$; (d) $P^+$ of sample $P^+/N$.

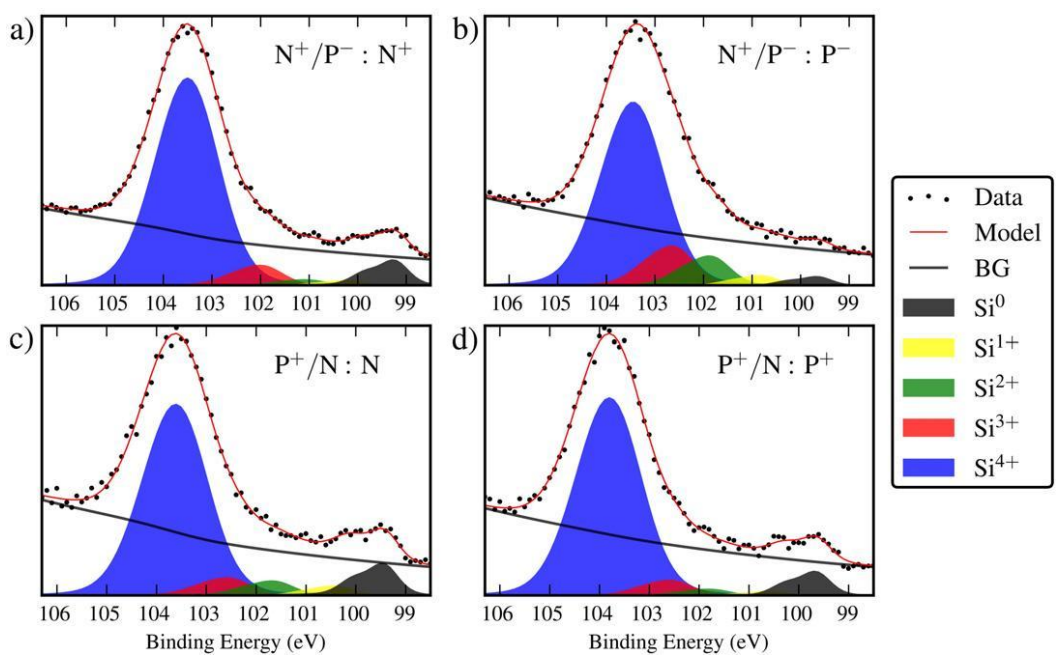

Fig. 5. Best fit to the Si 2p spectra extracted from a single pixel: (a) $N^+$ of sample $N^+/P^-$; (b) $P^-$ of sample $N^+/P^-$; (c) N of sample $P^+/N$; (d) $P^+$ of sample $P^+/N$. The spectra are, of course, noisier than the average spectra of Fig. 4, but the iterative fitting procedure clearly allows extraction of chemical state information pixel by pixel.

The use of hemispherical analyzers in the imaging system naturally introduces energy dispersion in one direction of the FoV called non-isochromaticity. This has a simple quadratic dependence on position: b×(Position−c)² [16]. In general, separate experiments must be performed to measure the non-isochromaticity, however in our case it is possible to determine it with sufficient accuracy from the same dataset. To perform this measurement, we adjust a parabola to the energy position of $Si^{4+}$ after a preliminary fit to the full model, in an area with constant doping, so that any energy shift can be attributed to the residual α²-aberration of the imaging spectrometer (Fig. 2). With this model of the non-isochromaticity we construct a map of the energy shift that we apply to all components of the model. The average value of these parameters and their standard deviation for sample $N^+/P^-$ are b= (−7.53±0.30)×10$^{-5}$ eV/ μm and c=(10.38±0.09) μm.

Another issue for quantitative microscopy is the non-uniform illumination (or transmission function) across the FoV. This can be caused by a vignetting effect on the lens coupling the scintillator and the image detection system (CCD) or by variation of the electron optics transmission function over the FoV (at constant kinetic energy). Transmission non-uniformity is measured using an image presenting homogeneous photoelectron emission. A first approximation is obtained using the average of the intensity of the last three images in the high KE end of the SI. After fitting, we use the value of the overall curve fitting model at the highest kinetic energy. The result, shown in Fig. 3, is free from any sample-related contrast and is considered to be a good estimation of the transmission properties of the imaging system as a whole (electron and light image transmission). The asymmetry in the electron transmission function is possibly linked to the offset observed in the value of the c parameter as obtained from the non-isochromaticity data of Fig. 2, which in a perfectly aligned optical system should of course indicate the central (binned) CCD pixel. The figure shows that the transmission can be as low as 50% of its maximum, showing the importance of this correction. The fit results were scaled according to this map.

The fitting procedure was performed as follows. To find suitable starting values for each pixel, we first identify differently doped regions by fitting only the most prominent feature of the spectra, the $Si^{4+}$ peak. We average the SI over the homogeneously doped regions far from the interface and the full model is fitted to the resulting spectra (Fig. 4). All the parameters of the model were free to float at this stage, except the Gaussian FWHM that were fixed to 0.28, 0.44, 0.58, 0.66 and 1.15 eV for $Si^0$, $Si^{1+}$, $Si^{2+}$, $Si^{3+}$ and $Si^{4+}$, respectively [4]. The isochromaticity was corrected by aligning the SI according to the non-isochromaticity map described above. The total spectral resolution is taken into account by convoluting the Voigt functions with a Gaussian representing the instrumental broadening and whose FWHM was free for this fit. The latter converges to 0.5 eV, close to the theoretical energy resolution. The Lorentzian part of the Voigt function represents the lifetime broadening and was found to be 0.2 eV FWHM. The values of the parameters obtained by fitting the model to the averaged spectra are later on used as starting parameters to the pixel-by-pixel fit in the FoV. To ensure a good fit in all the pixels the chi-squared map is checked after the fit of the full SI and, where required, the fit is repeated manually resetting the starting parameters until a good fit is reached. Fig. 5 illustrates the fit of the model to individual pixels in the differently doped regions.

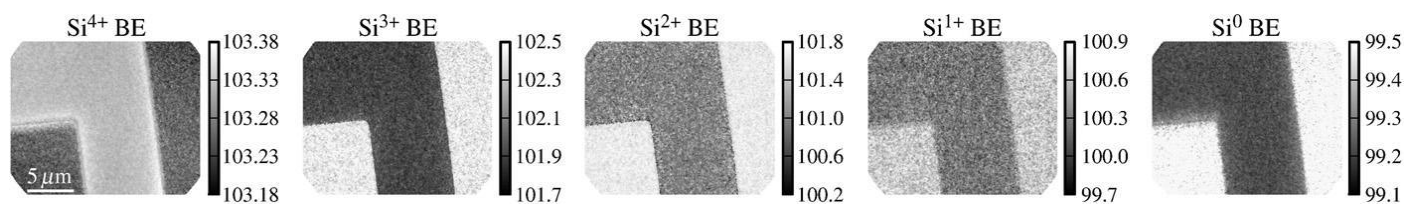

Fig. 6. Binding energy maps for $Si^0$ to $Si^{4+}$ obtained from the model based fit to the Si 2p spectra of sample $N^+/P^-$. Energy scales are in eV.

Table 2
Mean binding energies and standard deviations in eV for both $N^+/P^-$ and $P^+/N$ samples.

| Layer | $N^+/P^-$ | | $P^+/N$ | |
|---|---|---|---|---|
| | $N^+$ | $P^-$ | N | $P^+$ |
| $Si^0$ | 99.24±0.03 | 99.59±0.09 | 99.47±0.03 | 99.57±0.03 |
| $Si^{1+}-Si^0$ | 1.05±0.15 | 1.08±0.17 | 1.05±0.29 | 1.05±0.32 |
| $Si^{2+}-Si^0$ | 1.82±0.12 | 2.12±0.12 | 2.13±0.24 | 2.02±0.24 |
| $Si^{3+}-Si^0$ | 2.70±0.06 | 2.87±0.12 | 2.95±0.16 | 2.78±0.13 |
| $Si^{4+}-Si^0$ | 4.10±0.01 | 3.67±0.09 | 4.00±0.04 | 4.01±0.04 |

A weighted least squares algorithm was used for the fitting procedure. The weighting takes into account the Poissonian nature of the noise, and is appropriately scaled to account for the corrections applied to the original dataset. All the parameters (i.e. binding energy, intensity and background of each component) were free to float except the FWHM of the Gaussian and Lorentzian of the Voigt profiles and the spectral resolution that was fixed by the preliminary fit. The data analysis was performed with HyperSpy [27,28], a software suite originally developed for the analysis of EELS spectrum images obtained in transmission electron microscopy. HyperSpy is written in Python, an increasingly popular scripting language in the scientific community, which highly facilitated its adaptation for the present work. The least squares optimizer used was the Python wrapper around MINPACK's lmdif and lmder algorithms [29] (a modification of the Levenberg–Marquardt algorithm) included in the Scipy scientific library [30].

## 4. Results

### 4.1. Si 2p core-level analysis

We now present in detail the binding energy and intensity maps resulting from the pixel-by-pixel curve fitting analysis of the samples $N^+/P^-$ and $P^+/N$.

In both samples, the parameters of the background function, A and τ for the tail of the SSE and s for the Shirley background, are approximately constant for a given doping level. This is a first indication of the homogeneity of each pattern on the sample and the consistency of the analysis. However, going from $P^-$ to $N^+$ regions, both A and τ are higher. Given that the general form of the secondary electron peak must be the same whatever the doping and there are no significant direct transitions at ~20 eV from threshold, the difference can only be ascribed to an energy difference in threshold, in other words, to a shift in the work function. This qualitatively follows the expected change in the measured Fermi level position in the gap as a function of doping and surface photovoltage. On the other hand, the change in the Shirley factor as a function of the doping reflects a change in the loss spectra, and is more likely to be linked to the detailed oxide and sub-oxide structure although its interpretation is out with the scope of this work.

#### 4.1.1. $N^+$ patterns on $P^-$ substrate

Fig. 6 shows the binding energy maps for the substrate ($Si^0$) and $Si^{1+}$, $Si^{2+}$, $Si^{3+}$ and $Si^{4+}$, as obtained from the above described fit. The central region is the heavily doped $N^+$ zone and the surrounding areas correspond to light $P^-$ substrate doping. Within a given doped pattern the binding energy is constant apart from in the vicinity of a p–n junction, where there are evident changes corresponding to the band line-up. Furthermore, it is an indication of the accuracy of the non-isochromaticity correction since large areas show the same energy across the FoV. In this sense it is illustrative to compare the $Si^{4+}$ binding energy of Fig. 6 with that of Fig. 2 (a).

If one were to consider only the flat band scheme with Fermi level pinning as a function of doping, then the right hand map of Fig. 6, which is the $Si^0$ or substrate emission, is the contrary of what one would expect: highly N-doped Si should pin the Fermi level just below the conduction band minimum. However, as already shown [12], the value of the IMFP implies that the 2p electrons come from bent bands at the oxide interface, and it is the band bending which significantly modifies the band line-up. In itself this is not new; what is remarkable is the extent to which the band bending seems spatially homogeneous. Not only does this attest to the quality of the ion implantation, but it also validates the extraction of precise values for the band alignments. The average values of the BE for each chemical state are given in Table 2.

Fig. 7 shows the $Si^0$, $Si^{1+}$, $Si^{2+}$, $Si^{3+}$, and $Si^{4+}$ component intensity maps. The global information obtained from the fit to the average spectra of Fig. 4 (a) and (b) is confirmed, but there is considerably more quantitative information. We note that the intensities are uniform within a given doping pattern. This is evidence not only for homogeneous ion implantation, as expected, but also for the accurate correction of the transmission function in the FoV, which would otherwise cause a signal decrease of about 50% from the centre to the edges of the image (see Fig. 3). The chemically resolved Si 2p intensity maps will be used as input parameters to a 5 layer model of the $SiO_2$/Si interface in Section 5.

#### 4.1.2. $P^+$ patterns on N substrate

Fig. 8 shows the binding energy maps for the $Si^0$, $Si^{1+}$, $Si^{2+}$, $Si^{3+}$ and $Si^{4+}$ components. Within a given doping pattern, the binding energy is again constant. Moreover, sub-oxide binding energies show little or no BE contrast as a function of doping, in contrast to the sub-oxide binding energy maps of the $N^+/P^-$ sample of Fig. 5. Clearly the band line-up is different at the oxide/silicon interface for the two samples. The average BE values of each component are given in Table 2.

Fig. 9 shows the Si 2p intensity maps for the $Si^0$, $Si^{1+}$, $Si^{2+}$, $Si^{3+}$ and $Si^{4+}$. The contrast between N-doped and P-doped regions is clearly different from that obtained on the $N^+/P^-$ sample shown in Fig. 9. The high contrast observed for the $Si^{3+}$, $Si^{2+}$ and $Si^{1+}$ sub-oxide intensities on the $N^+/P^-$ sample is inversed and considerably attenuated on the $P^+/N$ sample.

Table 2 shows the average and standard deviation values for the binding energies of each chemical state in each doping level. The binding energy shifts from intrinsic Si to silicon oxide are in good agreement with literature values [4]. We note that the standard deviations for the ensemble of pixel-by-pixel spectra are very small and systematically smaller than the overall energy resolution. As a free parameter in our fitting procedure, this confirms the accuracy of the results. The correlation of the binding energy results with band bending requires further analysis as shown in [12] and discussed below. However, in this paper we focus on the structural information provided by the thickness model.

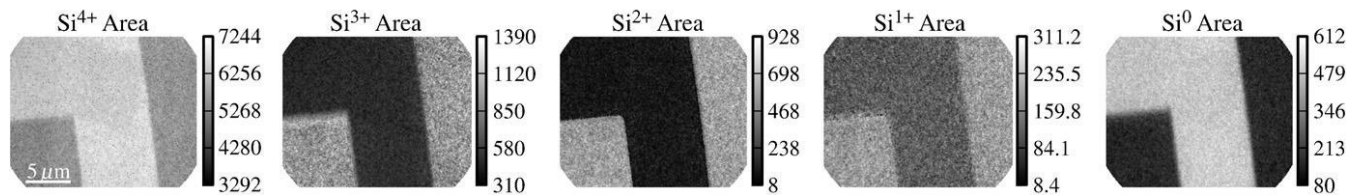

Fig. 7. $Si^0$, $Si^{1+}$, $Si^{2+}$, $Si^{3+}$ and $Si^{4+}$ peak area maps obtained from best fits to the Si 2p spectra generated from the $N^+/P^-$ SI.

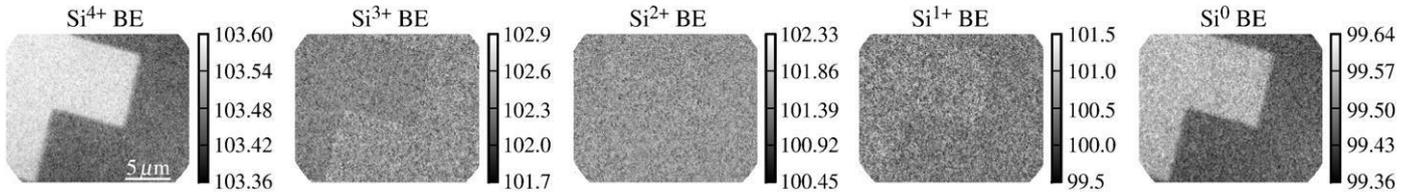

Fig. 8. Binding energy maps for $Si^0$ to $Si^{4+}$ obtained from the model based fit to the Si 2p spectra of sample $P^+/N$. Scale is in eV.

## 5. Discussion

The main focus of this paper is the spatially resolved quantification of the sub-oxide thicknesses as a function of the doping level and type of the micron scale silicon patterns.

However, a wealth of information is also available from the binding energy distribution as a function of doping. This has already been discussed. Here, we briefly recall the main conclusions [12]. The binding energy of a core electron in a doped semiconductor as measured by photoemission using an intense X-ray source can be written as:

$$BE = BE_{fb}^i + \Delta E_F^{doping} + \Delta E_{BB} + \Delta E_{SPV}, \quad (1)$$

where $BE_{fb}^i$, the binding energy, is measured in the intrinsic case and in flat band conditions; $E_F^{doping}$ is the shift in the position of the Fermi level within the gap as a function of doping with respect to intrinsic silicon; $E_{BB}$ is the band bending at the substrate/oxide interface; and $E_{SPV}$ is the energy shift due to the surface photovoltage (SPV), which is dependent on both the incident photon flux and the substrate electronic structure.

For example, for $N^+/P^-$ the measured binding energy is 99.3 eV in the heavily n-doped region, and 99.60 eV in the p-doped substrate. The effect of heavy n doping shifts the position of the Fermi level in the gap. With the low escape depth of the 2p electrons stimulated by the 127 eV incident photons, this is countered by the upwards band bending, which is itself partially flattened by the X-ray induced surface photovoltage. From the calculated SPV values we were able to estimate the real band bending at the oxide/substrate interface [12].

Now we turn to the extraction of the sub-oxide thicknesses as a function of position in the microscope FoV. We can obtain qualitative information about the oxides from the rich structure of the intensity maps of the Si core-level components (Figs. 7 and 9). For both samples the $Si^0$ emission from the n-type region is higher than from the p-type region, indicating a thicker oxide layer for the latter. There is high contrast in the $Si^{2+}$ valence state for the $N^+/P^-$ sample, whereas, for the $P^+/N$ sample the contrast is much lower. The $Si^{4+}$ emission also shows high contrast for both samples. Thus, both the sub-oxide structure and layer thicknesses depend on the doping. The presence of all of the Si sub-oxides indicates that the interface is far from abrupt. Here we have evidence of an interface whose extent depends on both the configuration ($N^+/P^-$ or $P^+/N$) and the doping type and level (n or p).

A quantitative analysis of the extent of the different oxidation states therefore requires a detailed model of the sub-oxide structure. A two-layer model was first proposed by Himpsel et al. [4] and has since been widely used for two layers. It has been successfully extended to three layers to simulate more complex gate oxide stacks [31]. We use a five-layer model for the sample (Fig. 10) to calculate the oxide layer thickness using core-level intensities. The starting point is to assume that the interface can be represented as a stack of five successive layers, $SiO_2/Si^{3+}/Si^{2+}/Si^{1+}/Si$, thus the nearer to the Si substrate, the lower the Si oxidation state. This assumption agrees with recent angle-resolved Si 2p photoelectron spectra results showing that the sub-oxides are distributed as a function of valence between the Si substrate and the $SiO_2$. [32]. A five-peak model has also been used by Seah [1]. However, in the latter, all of the sub-oxides were assumed to be at the same depth in the oxide stack. This is suitable for an accurate determination of the overlying $SiO_2$ thickness, however, we have favoured the successive stack geometry in order to examine the sub-oxide layer structure. In both cases, the resulting $SiO_2$ should be identical. Here, the major innovation is the application to an entire SI, in order to characterize the spatially resolved sub-oxide structure. In other words, the final model is adjusted independently with respect to 80 thousand photoemission spectra extracted from the 25μm FoV.

In the framework of the five-layer model we can write the intensity of the signal of each layer as follows:

$$I_{Si^i} = I_{Si^i}^\infty \left(1 - A_{Si^i}\right) \prod_{j=i+1}^{4} A_{Si^j} \quad i=0\ldots4$$

where Si indicates the layer of i oxidation state. $A_{Layer} = e^{-\frac{d_{Layer}}{\lambda_{Layer}}}$ is the attenuation of the electrons in the layer due to an IMFP $\lambda_{Layer}$, $d_{Layer}$ is the thickness and $I_{Layer}^\infty$ is the intensity from semi-infinite blocks with the same composition as the layer. This intensity is given by:

$$I_{Layer}^\infty = n_{Layer} \sigma_{Layer} \lambda_{Layer}$$

$n_{Layer}$ being the Si atomic density and $\sigma_{Layer}$ the photo-ionization cross-section.

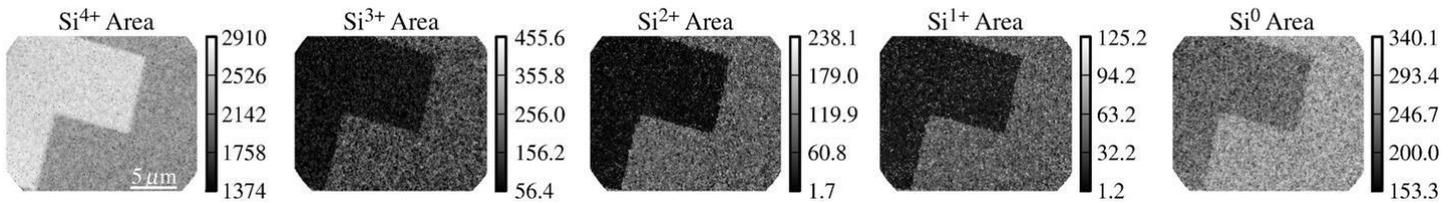

Fig. 9. $Si^0$, $Si^{1+}$, $Si^{2+}$, $Si^{3+}$ and $Si^{4+}$ peak area maps obtained from best fits to the Si 2p spectra generated from the $P^+/N$ SI.

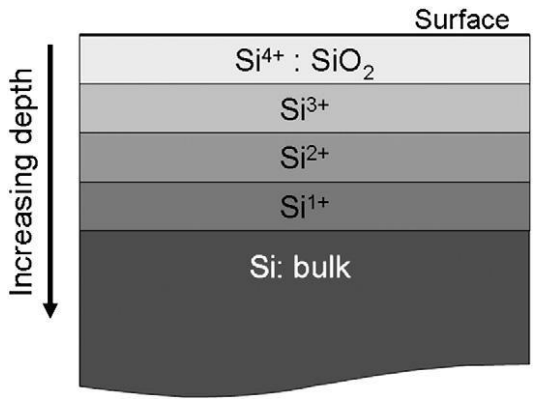

Fig. 10. Schematic of the five-layer model: each layer corresponds to a different chemical state of silicon, going from the substrate (bulk silicon, $Si^0$), through the sub-oxides ($Si^{1+}$, $Si^{2+}$, and $Si^{3+}$) to the surface (silica, $Si^{4+}$).

Dividing the equations for $1 \leq i \leq 4$ by the substrate signal intensity $I_{Si^0}$, we can obtain a set of 4 equations from which the thickness d of each layer can be calculated recursively:

$$A_{Si^i}^{-1} = \left(\frac{I_{Si^i}^{\infty}}{I_{Si^0}^{\infty}}\right)^{-1} \left(\frac{I_{Si^i}}{I_{Si^0}}\right) \prod_{j=1}^{i-1} A_{Si^j} + 1 \quad i = 1\ldots 4.$$

For each silicon oxidation state the ratio of the semi-infinite intensities to that of $Si^0$ at $h\nu = 127$ eV photon energy is required. Each of these ratios is the product of three ratios: the Si atomic densities $n_{Layer}/n_{Si}$, the IMFPs $\lambda_{Layer}/\lambda_{Si}$ and the photo-ionization cross-sections $\sigma_{Layer}/\sigma_{Si}$. Following Himpsel [4], there is a resonance near 130 eV for $Si^{3+}$ and $Si^{4+}$, we have therefore interpolated the cross-section values given for 120 and 130 eV, and assumed that the $Si^{3+}$ cross-section scales in the same way as $Si^{4+}$, obtaining $\sigma_{SiO_2}/\sigma_{Si} = 2.08$ and $\sigma_{Si^{3+}}/\sigma_{Si} = 1.6$. The atomic silicon concentration in $SiO_2$ is $n_{SiO_2} = 2.2 \times 10^{22}$ at/cm$^{-3}$ and in silicon $n_{Si} = 5.0 \times 10^{22}$ at/cm$^{-3}$. For the intermediate oxides we assume that the concentration scales with the valence state. We use an IMFP of 0.7 nm for silicon oxide and 0.33 nm for the Si substrate [31], the IMFPs in the sub-oxide layers are interpolated linearly with the oxidation state. The parameters are given in Table 3.

Using this model, the thickness of the $SiO_2$ and each sub-oxide layer can be calculated from the intensities associated with the respective components in the Si 2p spectrum at each pixel of the whole dataset. Hence, oxide and sub-oxide thickness maps can be obtained as shown in Fig. 11 (a) and (b).

To illustrate thickness variations across the different doping levels Fig. 12 shows the oxide and sub-oxide thickness profiles along a line of length 6.1 μm perpendicular to the p–n junctions. An alternative presentation is shown in Fig. 13 which stacks emphasizing the topography of the buried interface.

The oxide layer is always thicker over the p-doped regions than over the n-doped regions. Secondly, it is thicker on the $N^+/P^-$ sample than on the $P^+/N$ sample, whatever the doping. In the former, over the heavily n-doped regions the native oxide has a total thickness of 1.76 nm, of which 1.34 nm is due to $SiO_2$.

Table 3
Estimated photoemission parameters for Si 2p electrons at 127 eV photon energy, used in the model of the Si 2p core-level component intensities.

| Layer | $\sigma_{Layer}/\sigma_{Si}$ | $\lambda_{Layer}/\lambda_{Si}$ | $n_{Layer}/n_{Si}$ | $I^{\infty}_{Layer}/I^{\infty}_{Si}$ |
|---|---|---|---|---|
| $Si^{1+}$ | 1 | 1.28 | 0.86 | 1.10 |
| $Si^{2+}$ | 1.10 | 1.56 | 0.73 | 1.25 |
| $Si^{3+}$ | 1.60 | 1.84 | 0.59 | 1.74 |
| $Si^{4+}$ | 2.08 | 2.12 | 0.46 | 2.03 |

The total sub-oxide thickness is 0.42 nm, three quarters of which can be attributed to the $Si^{3+}$ rich layer. Therefore the interface, mainly composed of $Si^{1+}$ and $Si^{2+}$, appears sharp. In contrast, the oxide over the $P^-$ doped regions is thicker, 2.19 nm. The total sub-oxide thickness over the $P^-$ regions is almost three times that over the $N^+$ zones and there are marked differences in the thicknesses of each sub-oxide as a function of doping. The $SiO_2$ decreases by 0.26 nm whereas all of the sub-oxide thicknesses increase. The most spectacular change is that of the $Si^{2+}$ layer, which increases by 0.43 nm. Since Si–O bond lengths are typically 1.6–1.8 Å, the extended interface associated with Si in the 2+ oxidation state must be two to three atomic layers thick over the $P^-$ pattern. The presence of a distinct $Si^{2+}$ sub-oxide layer over Si(100) was also reported by Rochet et al. [5], although much thinner, probably because they grew a thermal oxide in vacuum, rather than studying the native oxide as in our case. In contrast, the $P^+/N$ sample presents a total oxide thickness of 1.49(1.64) nm over the $N(P^+)$ regions, mainly due to the $SiO_2$ layer, whereas the sub-oxide thicknesses are almost constant.

The thicknesses of each zone and layer are best summarized by considering the mean values and their standard deviations in each differently doped region far from the interface. The results are shown in Table 4. The standard deviation for the measured thickness is significantly smaller than typical Si–O bond lengths, which suggests that the model of five distinct layers is reasonable. The exception is the standard deviation in the $Si^{2+}$ thickness over the $N^+$ region of the $N^+/P^-$ sample which is about the same as the thickness obtained for the $Si^{2+}$ layer of the $P^+/N$ sample. This is partly due to the low intensity of the sub-oxide component, however, other low intensity components give smaller standard deviations. Thus, the $Si^{2+}$ layer probably has significant atomic roughness and a high concentration of defects. The expected variations in silicon–oxygen bond lengths are much smaller, for example, Giustino and Pasquarello [33] showed that the Si–O bond length could only vary by up to ±0.005 nm.

Across the p–n junctions in both samples, the spatial variation of the thickness of the different layers is smooth, which is an extra indication of the quality of the analysis. By fitting an error function to the curve of the Si0/Si+ interface for both samples we can estimate the spatial extent of the thickness change across the junction. The presentation in Fig. 13 shows a slightly wider depletion region in the $P^+/N$ sample.

The width over which the sub-oxide structure changes at the p–n junction is smaller in the case of the $N^+/P^-$ sample than the expected depletion width, whereas in the case of the $P^+/N$ sample it is larger. One reason may be that the width as measured here depends on the statistics associated with the $Si^{1+}$ emission, which is considerably attenuated, however, this is to a large extent compensated by the full field analysis, as can be inferred from Figs. 4 and 5. The reason might be intrinsic to p–n junctions. The band alignment will generate a local lateral electric field which can considerably alter the trajectories of the photoelectrons [34]. The field will be directed from the n- to p-type regions; however its actual magnitude will depend both on the Fermi level position and the depletion width, which are clearly different for the two samples. For a p–n junction between heavily doped n-type silicon and lightly doped p-type silicon, the band offset is about 1 eV and the depletion width of the order of 1 μm. This will induce a lateral electric field of ~1 kV/mm, significant with respect to the 7 kV/mm of the extractor lens. Thus, in the region of the junction the photoelectron trajectories will be bent by the lateral electric field, and the junction position and even width in the PEEM images may not correspond to the real values. We have tried several extractor lens voltage settings in order to vary the nominal lateral resolution between 150 and 70 nm. At threshold, i.e. for very low kinetic energies, we were unable to improve the measured resolution across the junction below 105 nm. Thus, we believe that the lateral electric field in the vicinity of the junction limits the resolution to 100 nm.

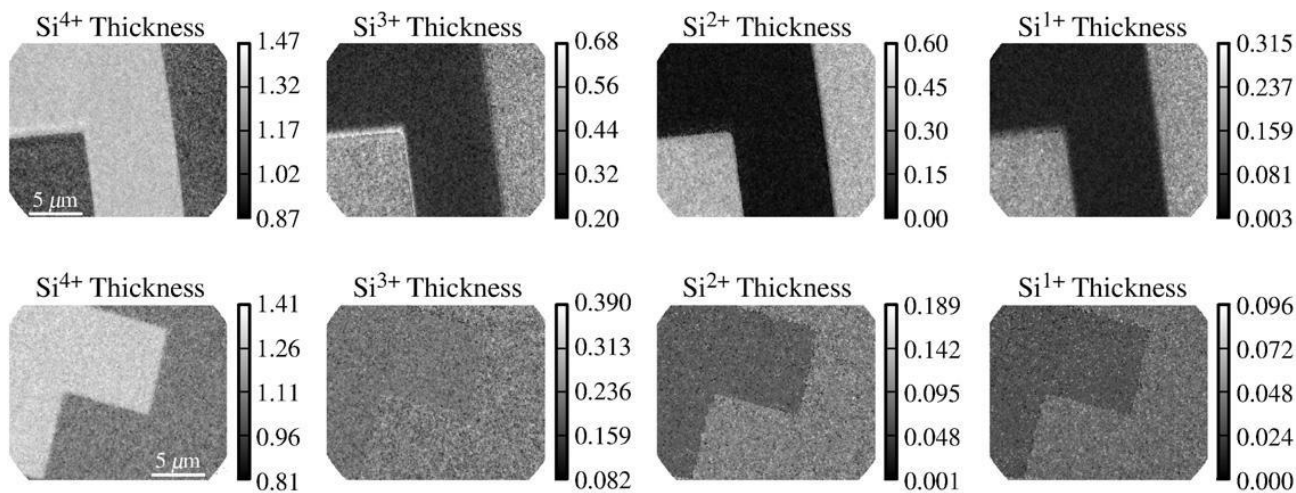

Fig. 11. Oxide and sub-oxide thickness maps for (a) sample N$^+$/P$^-$ and (b) P$^+$/N derived from the five-layer model of Fig. 10. Thickness scale in nm.

For the higher KE Si 2p electrons, this figure should represent an upper limit. We have checked the surface topography of the native oxide using AFM. The doped patterns are systematically higher than their respective substrates. The height differences are 1 nm for the N$^+$/P$^-$ sample and 0.5 nm for the P$^+$/N sample. These differences are introduced during ion implantation. They do not change the sub-oxide thicknesses as deduced from photoemission. The structure visible in the Si$^{4+}$ and Si$^{3+}$ thickness maps in Fig. 11 (a) is possibly due to such topography. However, to incorporate the AFM topography into Fig. 13 requires knowledge of the positional shifts due to the lateral electric field across the junction. Once this is done, it will be possible to describe the full surface and sub-surface topographies by a combined and non-destructive XPEEM–AFM analysis. The effect of the lateral electric field is therefore of general relevance to PEEM imaging of semiconductor structures and is currently under study.

Two other possible effects in photoemission need to be considered. Exposure for several hours to high intensity soft X-ray radiation can induce room temperature desorption of semiconductor oxides. Heun et al. demonstrated this in the case of SiO$_2$ [35]. However the brilliance of the X-ray source used by them was three to four orders of magnitude higher than that on the CIPO beamline used in these experiments. Thus we can exclude radiation induced thinning of the native oxide in our results. A comparative study of the oxide formation on Si(100) for equivalent B or P doping profiles has shown a faster SiO$_2$ growth on n-doped silicon than on p-doped silicon [36]. This is what we observe for the N$^+$/P$^-$ sample. On the other hand, for the P$^+$/N sample we observe a thicker SiO$_2$ layer over the P$^+$ doped region. As explained above, the latter sample required successive P and B implantation, thus the regions cannot be simply compared to macroscopic doped silicon wafers. Using Vegard's law as a first approximation and assuming that the post implantation annealing was sufficient to produce a homogeneously doped layer we can estimate the strain induced in the Si substrate by doping [37], however, the resulting isotropic strains for the two samples are less than 0.01%. The differences in the oxide growth are therefore more likely to be due to electrical enhancement of the oxidation. Finally, charging of SiO$_2$ overlayers is also known to affect the measured binding energy, however, this only becomes significant for oxides thicker than about 2 nm [38]. Furthermore, we have measured no time dependence of the binding energy, suggesting that charging in these films is negligible.

## 6. Conclusion

We have used full field energy-filtered photoelectron emission microscopy to analyze the spatial variation of the binding energies and thicknesses of the native oxide layers on p- and n-doped micron scale silicon patterns. Using a pixel-by-pixel spectral analysis and by correcting for non-isochromaticity, microscope transmission functions and synchrotron photon flux decay, quantitative information is made available. From a five-layer model for the Si 2p core-level intensities we can deduce the variations in the extent of the different oxide layers. Both samples are thicker in the p-doped areas, the N$^+$/P$^-$ sample presenting a total variation in thickness two and a half times that of the $P^+/N$ sample. Furthermore, the interface oxides show a marked thickness variation in the former while in the latter the interface thickness stays constant. On the other hand, the relative differences in the thickness of the overlying $SiO_2$ are lower, suggesting that the extension of the interface sub-oxide layer is linked to the doping-dependent surface electronic band alignments.

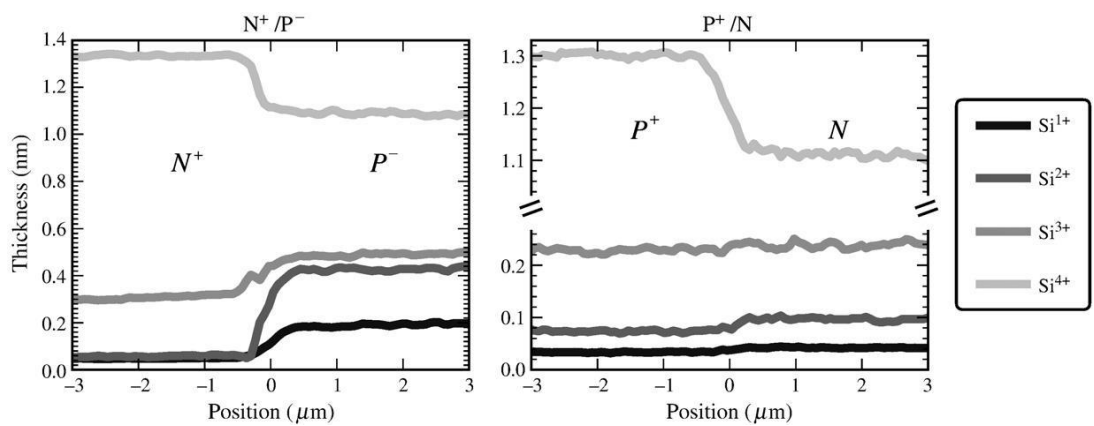

Fig. 12. Thickness profiles for each silicon oxidation state. In this figure, the profiles display the thickness of each layer independently.

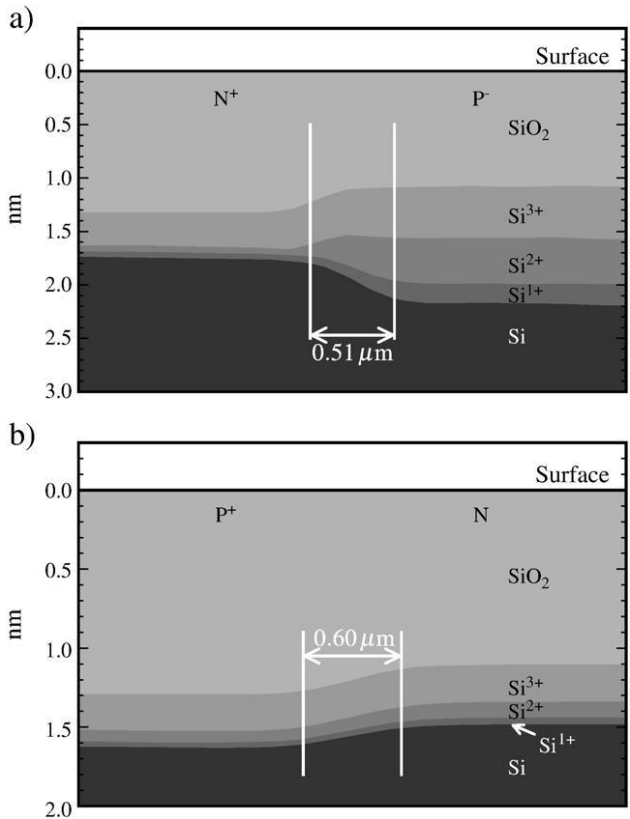

Fig. 13. Spatial distribution of the $SiO_2$ and Si sub-oxide thickness distribution across a p–n junction as deduced from a five-layer model adjusted using the pixel-by-pixel Si 2p core-level intensities. (a) $N^+/P^-$. (b) $P^+/N$. In this figure, the profiles are summed to illustrate the real profile across the interface. The topography determined by Atomic Force microscopy (AFM) is not included, see below. Please note the different scales on the y axes. An error function is fitted to the profile of the interface between Si and $Si^{1+}$ to measure the spatial extent of the interface, the value of $2\sigma$ is given in the figure.

## Acknowledgements

We would like to thank N. Zema, S. Turchini, and B. Delomez for their technical help during the measurements on the CIPO beamline at ELETTRA (Trieste, Italy) and D. Mariolle for the AFM measurements. This work has been supported by the French National Research Agency (ANR) through the "Recherche Technologique de Base" Program and the project 05-NANO-065 XPEEM. F. de la Peña acknowledges support by the European Community Marie Curie Action (No. MEST-CT-2004-514307). M. Walls and F. de la Peña acknowledge support from the ESTEEM IP3 project within the 6th Framework Programme of the European Commission.

Table 4
Mean value and standard deviations for oxide and sub-oxide layer thicknesses in nm for $N^+/P^-$ and $P^+/N$ samples.

| Layer | $N^+/P^-$ | | $P^+/N$ | |
|---|---|---|---|---|
| | $N^+$ | $P^-$ | N | $P^+$ |
| $Si^0$ | – | – | – | – |
| $Si^{1+}$ | 0.05±0.02 | 0.19±0.03 | 0.04±0.01 | 0.03±0.01 |
| $Si^{2+}$ | 0.06±0.03 | 0.49±0.06 | 0.10±0.02 | 0.07±0.02 |
| $Si^{3+}$ | 0.31±0.06 | 0.43±0.05 | 0.24±0.04 | 0.23±0.03 |
| $Si^{4+}$ | 1.34±0.03 | 1.08±0.06 | 1.11±0.05 | 1.31±0.04 |